\def\be{\begin{equation}}
\def\ee{\end{equation}}
\def\ba{\begin{array}}
\def\ea{\end{array}}

\documentclass[11pt]{article}
 \usepackage{graphicx}

\input amssym.def
\begin{document}
\baselineskip=18pt \setcounter{page}{1}
\begin{center}
{\LARGE \bf Three-tangle for high-rank mixed states}
\end{center}
\vskip 2mm

\begin{center}
{\normalsize Shu-Juan He$^1$, Xiao-Hong Wang$^2$,
Shao-Ming Fei$^{2}$, Hong-Xiang Sun$^3$ and Qiao-Yan
Wen$^1$}
 
\medskip

\begin{minipage}{4.8in}
{\small \sl $ ^1$ State Key Laboratory of Networking and Switching Technology, Beijing University of Post and
Telecommunication, China}\\
{\small \sl $ ^2$ Department of Mathematics, Capital  Normal
University, Beijing,
China}\\
%{\small \sl $ ^3$ Institute of Applied Mathematics, University of
%Bonn,  53115 Bonn, Germany}\\
{\small \sl$ ^3$ School of Science, Beijing University of Post and
Telecommunication, China}

\end{minipage}
\end{center}

\vskip 2mm
\begin{center}

\begin{minipage}{4.8in}

\centerline{\bf Abstract}
\bigskip

A family of rank-n ($n=5,6,7,8$) three-qubit mixed states are
constructed. The explicit expressions for the three-tangle and
optimal decompositions for all these states are given. The CKW
relations for these states are also discussed.

\vskip 9mm
%{\sl \noindent AMS Subject Classification:}
 \vskip 1mm PACS
number(s): 03.67.Mn, 03.65.Ud
\end{minipage}
\end{center}

\vskip0.4cm

\section{Introduction }

Quantum entangled states are important physical resource
and play key roles in quantum information processing such as
teleportation, superdense coding, quantum cloning, quantum
cryptography \cite{information,teleportation,cloning}.
Characterizing and quantifying entanglement of quantum
states are of great importance.

To quantify entanglement some measures like concurrence
\cite{Rungta01-AlbeverioFei01}, entanglement of formation
\cite{BDSW,Horo-Bruss-Plenioreviews} are studied. Though
entanglement of bipartite states have been understood well in many
aspects \cite{two-qubit}, there is still no generally accepted
theory for characterizing and quantifying entanglement for
multipartite qubit systems, especially for mixed states. For
three-qubit systems, some results have been presented
\cite{oufan,gfw,gf,cw}. An important quantity for three-qubit
entanglement is the so called residual entanglement or three-tangle
\cite{ckw-relation}, which is a polynomial invariant for three-qubit
states, the modulus of the hyperdeterminant \cite{Caley,Miyake}.

For a pure three-qubit state
$|\psi\rangle=\sum_{i,j,k=0}^{1}a_{ijk}|ijk\rangle\in {\cal C}^2 \otimes {\cal
C}^2 \otimes {\cal C}^2$, its three-tangle is defined by
\be\label{three-tangle-pure}
\tau_{3}(|\psi\rangle)=4|d_{1}-2d_{2}+4d_{3}|,
\ee
where
$$d_{1}=a_{000}^{2}a_{111}^{2}+a_{001}^2a_{110}^2+a_{010}^{2}a_{101}^{2}+a_{100}^{2}a_{011}^{2},$$
$$d_{2}=a_{000}a_{111}a_{011}a_{100}+a_{000}a_{111}a_{101}a_{010}+a_{000}a_{111}a_{110}a_{001}$$
$$+a_{011}a_{100}a_{101}a_{010}+
a_{011}a_{100}a_{110}a_{001}+a_{101}a_{010}a_{110}a_{001},$$
$$d_{3}=a_{000}a_{110}a_{101}a_{011}+a_{111}a_{001}a_{010}a_{100}.$$
For a mixed three-qubit state $\rho=\sum_i p_i\rho_i$, $0<p_i\leq
1$, $\rho_i=|\psi_i\rangle\langle\psi_i|$, $|\psi_i\rangle\in{\cal
C}^2 \otimes {\cal C}^2 \otimes {\cal C}^2$, the three-tangle is
defined in terms of convex roof \cite{convex-roof} \be\label{mix}
\ba{lll}\tau_{3}(\rho)=min\sum_{i}p_{i}\tau_{3}(\rho_{i}).\ea \ee

A decomposition that realizes the above minimum is called optimal.
It is a challenge to find the optimal decomposition even for the
simplest case of rank-2 mixed states. Nice analytical results have
been obtained for some classes of three-qubit mixed states. In
\cite{rank2,grank2} Lohmayer et al. have constructed the optimal
decomposition for a family of rank-2 three-qubit states. Jung et
al.\cite{rank3,rank4} have also provided analytical formulae of
three-tangle for a class of rank-3 and rank-4 three-qubit mixed
states. In \cite{guogc} a numerical method has been also presented
to compute the three tangle for general three-qubit states.

In this paper we analyze the optimal decomposition for
some families of high rank-$n$ ($n=5,6,7,8$) three-qubit mixed states.
The analytical expressions for the three-tangles and explicit
optimal decompositions for all these states are given. The CKW
relations for these states are also investigated.

\section{The Three-tangle for some high-rank mixed states}

Recently, Jung et al.\cite{rank4} have provided an analytic quantification of the three-tangle for a rank-4 three-qubit mixed state which is composed by GHZ-type states. In this paper, we extend the method of \cite{rank4} to the high-rank mixed states.

We use the following notations \cite{rank4} in studying three tangle of
high rank mixed states:
$$\ba{l}
|GHZ,1\pm\rangle=\frac{1}{\sqrt{2}}(|000\rangle\pm|111\rangle),~~~
|GHZ,2\pm\rangle=\frac{1}{\sqrt{2}}(|110\rangle\pm|001\rangle),\\
|GHZ,3\pm\rangle=\frac{1}{\sqrt{2}}(|101\rangle\pm|010\rangle),~~~
|GHZ,4\pm\rangle=\frac{1}{\sqrt{2}}(|011\rangle\pm|100\rangle).
\ea
$$

{\sf The case of rank-5 states} We first consider the following
rank-5 states: \be\label{rank5}
\ba{lll}\sigma(p)=p|GHZ,1+\rangle\langle
GHZ,1+|+(1-p)\Gamma_{GHZ},\ea \ee where
$\Gamma_{GHZ}=\frac{1}{10}|GHZ,1-\rangle\langle
GHZ,1-|+\frac{3}{10}|GHZ,2+\rangle\langle GHZ,2+|
+\frac{3}{10}|GHZ,3+\rangle\langle
GHZ,3+|+\frac{3}{10}|GHZ,4+\rangle\langle GHZ,4+|$.

We first consider the state $\Gamma_{GHZ}$. In order to calculate
the three-tangle of the state $\Gamma_{GHZ}$, we first investigate
the following rank-4 state:
%\be\label{2} \ba{lll}
$$\rho(p)=p|GHZ,1-\rangle\langle
GHZ,1-|+(1-p)\Pi_{GHZ},$$%\ea \ee
where $\Pi_{GHZ}=\frac{1}{3}[|GHZ,2+\rangle\langle
GHZ,2+|+|GHZ,3+\rangle\langle GHZ,3+|+|GHZ,4+\rangle\langle
GHZ,4+|]$ which  has vanishing three-tangle \cite{rank4}.

By straightforward calculation the three tangle of the following pure state
$$|Z(p,\varphi_{1},\varphi_{2},\varphi_{3})\rangle
=\sqrt{p}|GHZ,1-\rangle-\sqrt{\frac{1-p}{3}}(e^{i\varphi_{1}}|GHZ,2+\rangle
+e^{i\varphi_{2}}|GHZ,3+\rangle+e^{i\varphi_{3}}|GHZ,4+\rangle)$$
is given by
\be\label{4}\ba{rcl}
\tau_{3}(|Z(p,\varphi_{1},\varphi_{2},\varphi_{3})\rangle)&=&
|p^2+\frac{(1-p)^2}{9}(e^{4i\varphi_{1}}+e^{4i\varphi_{2}}+e^{4i\varphi_{3}})
+\frac{2}{3}p(1-p)(e^{2i\varphi_{1}}+e^{2i\varphi_{2}}+e^{2i\varphi_{3}})\\[3mm]
&&-\frac{2(1-p)^2}{9}(e^{2i(\varphi_{1}+\varphi_{2})}+e^{2i(\varphi_{1}+\varphi_{3})}
+e^{2i(\varphi_{2}+\varphi_{3})})|.
\ea \ee

Note that $\tau_{3}(|Z(p,\varphi_{1},\varphi_{2},\varphi_{3})\rangle)$ is zero
at $\varphi_{1}=\varphi_{2}=\varphi_{3}=0$ and
$p_{0}=\frac{2-\sqrt{3}}{2}\doteq0.134.$ And $\rho(p)$ can be decomposed into $\rho(p)=\frac{p}{8p_{0}}\sum\Gamma_{i}(p_{0})+\frac{p_{0}-p}{p_{0}}\prod_{GHZ}$ for $0\leq p\leq p_{0}$, where
\be\label{def1}\ba{l}\Gamma_1(p_{0})=|Z(p_{0},0,0,0)\rangle\langle
Z(p_{0},0,0,0)|,~~~
\Gamma_2(p_{0})=|Z(p_{0},0,0,\pi)\rangle\langle Z(p_{0},0,0,\pi)|,\\[3mm]
\Gamma_3(p_{0})=|Z(p_{0},0,\pi,0)\rangle\langle Z(p_{0},0,\pi,0)|,~~~
\Gamma_4(p_{0})=|Z(p_{0},0,\pi,\pi)\rangle\langle Z(p_{0},0,\pi,\pi)|,\\[3mm]
\Gamma_5(p_{0})=|Z(p_{0},\pi,0,0)\rangle\langle Z(p_{0},\pi,0,0)|,~~~
\Gamma_6(p_{0})=|Z(p_{0},\pi,0,\pi)\rangle\langle Z(p_{0},\pi,0,\pi),\\[3mm]
\Gamma_7(p_{0})=|Z(p_{0},\pi,\pi,0)\rangle\langle Z(p_{0},\pi,\pi,0)|,~~~
\Gamma_8(p_{0})=|Z(p_{0},\pi,\pi,\pi)\rangle \langle
Z(p_{0},\pi,\pi,\pi)|.\ea\ee
All $\Gamma_{i}(p_{0})$ and $\prod_{GHZ}$'s three-tangle are zero, therefore the three-tangle for the mixed state $\rho(p)$ is zero for $0\leq p\leq
p_{0}$. Thus $\Gamma_{GHZ}$ has vanishing three-tangle.

Now consider the three-qubit pure state constituted by linear combinations of
$|GHZ,1+\rangle$, $|GHZ,1-\rangle$, $|GHZ,2+\rangle$, $|GHZ,3+\rangle$ and $|GHZ,4+\rangle$:
\be\label{5}\ba{rcl}
|Z(p,\varphi_{1},\varphi_{2},\varphi_{3},\varphi_{4})\rangle
&=&\sqrt{p}|GHZ,1+\rangle-e^{i\varphi_{1}}\sqrt{\frac{1-p}{10}}|GHZ,1-\rangle
-e^{i\varphi_{2}}\sqrt{\frac{3(1-p)}{10}}|GHZ,2+\rangle\\[3mm]
&&-e^{i\varphi_{3}}\sqrt{\frac{3(1-p)}{10}}|GHZ,3+\rangle-
e^{i\varphi_{4}}\sqrt{\frac{3(1-p)}{10}}|GHZ,4+\rangle.
\ea \ee
The corresponding three-tangle is
\be\label{6}\ba{rcl}
\tau_{3}(|Z(p,\varphi_{1},\varphi_{2},\varphi_{3},\varphi_{4})\rangle)
&=&|p^2+\frac{(1-p)^2}{100}e^{4i\varphi_{1}}+
\frac{9(1-p)^2}{100}(e^{4i\varphi_{2}}+e^{4i\varphi_{3}}+e^{4i\varphi_{4}})\\[3mm]
&&-\frac{1}{5}p(1-p)e^{2i\varphi_{1}}-\frac{3}{5}p(1-p)(e^{2i\varphi_{2}}+e^{2i\varphi_{3}}+e^{2i\varphi_{4}})\\[3mm]
&&+\frac{3(1-p)^2}{50}(e^{2i(\varphi_{1}+\varphi_{2})}+e^{2i(\varphi_{1}+\varphi_{3})}+e^{2i(\varphi_{1}+\varphi_{4})})
\\[3mm]
&&-\frac{9(1-p)^2}{50}(e^{2i(\varphi_{2}+\varphi_{3})}
+e^{2i(\varphi_{2}+\varphi_{4})}+e^{2i(\varphi_{3}+\varphi_{4})})\\[3mm]
&&-\frac{6}{25}\sqrt{30}\sqrt{p(1-p)^3}e^{i(\varphi_{2}+\varphi_{3}+\varphi_{4})}|.\ea
\ee

Since the three-tangle
$\tau_3(|Z(p,\varphi_{1},\varphi_{2},\varphi_{3},\varphi_{4})\rangle)=0$
at $p=p_{0}=0.7377$ and
$\varphi_{1}=\varphi_{2}=\varphi_{3}=\varphi_{4}=0$, the state
$\sigma(p)$ can be expressed in terms of
$|Z(p,\varphi_{1},\varphi_{2},\varphi_{3},\varphi_{4})\rangle$,
\be\label{optidec1} \ba{rcl} \sigma(p)&=&\frac{1}{8}\sum\Pi_{i}(p),
\ea \ee where
\be\label{def1}\ba{l}\Pi_1(p)=|Z(p,0,0,0,0)\rangle\langle
Z(p,0,0,0,0)|,~~~
\Pi_2(p)=|Z(p,0,0,\pi,\pi)\rangle\langle Z(p,0,0,\pi,\pi)|,\\[3mm]
\Pi_3(p)=|Z(p,0,\pi,0,\pi)\rangle\langle Z(p,0,\pi,0,\pi)|,~~~
\Pi_4(p)=|Z(p,0,\pi,\pi,0)\rangle\langle Z(p,0,\pi,\pi,0)|,\\[3mm]
\Pi_5(p)=|Z(p,\pi,0,0,0)\rangle\langle Z(p,\pi,0,0,0)|,~~~
\Pi_6(p)=|Z(p,\pi,0,\pi,\pi)\rangle\langle Z(p,\pi,0,\pi,\pi)|,\\[3mm]
\Pi_7(p)=|Z(p,\pi,\pi,0,\pi)\rangle\langle Z(p,\pi,\pi,0,\pi)|,~~~
\Pi_8(p)=|Z(p,\pi,\pi,\pi,0)\rangle \langle
Z(p,\pi,\pi,\pi,0)|.\ea\ee

When $0\leq p\leq p_{0}$, we have the optimal decomposition of
$\sigma(p)$: \be\label{optidec} \ba{rcl}
\sigma(p)&=&\frac{p}{8p_{0}}\sum\Pi_{i}(p_0)+\frac{p_{0}-p}{p_{0}}\Gamma_{GHZ},
\ea \ee where $\Pi_{i}(p)$ are defined as (\ref{def1}),
$i=1,2,\cdots,8$. Since all $\Pi_{i}(p_0)$ and $\Gamma_{GHZ}$ has vanishing three-tangle,
we have that $\tau_{3}(\sigma(p))=0$ when $0\leq p\leq p_{0}$.

For $p>p_{0}$, the decomposition in Eqs.(\ref{optidec1}) also is a
trial optimal decomposition for  $\sigma(p)$. Its three-tangle is
\be\label{tI}
\ba{lll}g_{I}(p)=p^2-2p(1-p)-\frac{2(1-p)^2}{25}-\frac{6\sqrt{30}}{25}\sqrt{p(1-p)^3},~~p>p_{0}.
\ea \ee We need to check whether the function $g_{I}(p)$ is convex
or not for $p>p_{0}$. It can be verified that the function
$g_{I}(p)$ is convex for $p<p_{*}=0.9750$, but concave for
$p>p_{*}$. For large $p$ let us propose a decomposition of
$\sigma(p)$ as follows: \be\label{optidec2}\ba{rcl}
\sigma(p)&=&\frac{1-p}{8(1-p_{1})}\sum\Pi_{i}(p_1)+
\frac{p-p_{1}}{1-p_{1}}|GHZ,1+\rangle\langle GHZ,1+|, \ea \ee where
$p_{1}\leq p\leq 1$, $p_{1}\leq p_{\ast}$, $\Pi_{i}(p)$ are defined
as (\ref{def1}), $i=1,2,\cdots,8$.

The three tangle of (\ref{optidec2}) is given by \be\label{tII}
\ba{lll}
g_{II}(p)=\frac{p-p_{1}}{1-p_{1}}+\frac{1-p}{1-p_{1}}g_{I}(p_{1}).
\ea \ee Since $d^{2}g_{\Pi}/dp^2=0$ for all $p$, from $\partial
g_{II}/\partial p_{1}=0$ we have
$$3\sqrt{30}p_{1}^{\frac{1}{2}}(1-p_{1})^{-\frac{1}{2}}-3
\sqrt{30}p_{1}^{-\frac{1}{2}}(1-p_{1})^{\frac{1}{2}}=73,$$
which gives rise to
$$p_{1}=\frac{1}{2}+\frac{73\sqrt{6409}}{12818}\doteq0.9559.$$
Therefore the three tangle of the rank-5 state $\sigma(p)$ is given by
\be\label{imc} \ba{lll}\tau_{3}(\sigma(p))=\left\{
                                           \begin{array}{ll}
                                             0, & \hbox{$0\leq p\leq p_{0}$,} \\
                                             g_{I}(p), & \hbox{$p_{0}\leq p\leq p_{1}$,} \\
                                             g_{II}(p), & \hbox{$p_{1}\leq p\leq 1$,}
                                           \end{array}
                                         \right.
\ea \ee
  where $p_{0}$=0.7377, $p_{1}$=0.9559,
$g_{I}(p)$ is given by (\ref{tI}) and $g_{II}(p)$ by (\ref{tII}).
And the corresponding optimal decomposition are (\ref{optidec}),
(\ref{optidec1}) and (\ref{optidec2}) respectively. In order to show that (\ref{imc})
is genuine optimal, we plot the p-dependence of the three-tangles for
various $\varphi_{1}$, $\varphi_{2}$, $\varphi_{3}$, $\varphi_{4}$. These curves
have been referred as the characteristic curves \cite{curve}. As \cite{curve} indicated,
the three-tangle is a convex hull of the minimum of the characterisitic curves.
Fig.1 indicates that the three-tangles plotted as black solid line are the convex
characteristic curves, which implies that (\ref{imc}) is really optimal.
%\begin{figure}[h]
%\begin{center}
%\resizebox{8cm}{!}{\includegraphics{juan.eps}}
%\caption{The\$p$-dependence of one-tangle (upper dotted line), sum
%of the squared concurrences (solid line along the horizontal axis)
%and three-tangle (solid line)}.{\label{33}}
%\end{center}
%\end{figure}
\begin{figure}[h]
\begin{center}
\resizebox{8cm}{!}{\includegraphics{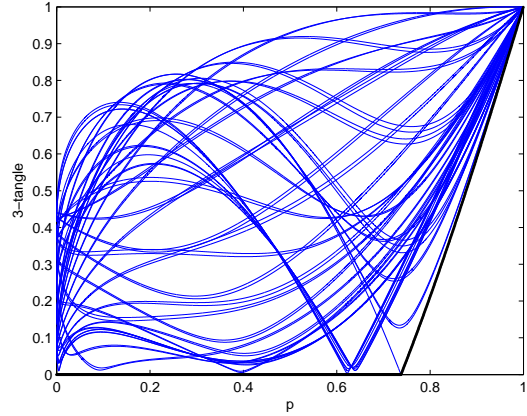}} \caption{The\ $p$
dependence of various $\varphi_{1}$, $\varphi_{2}$, $\varphi_{3}$, $\varphi_{4}$. We have chosen $\varphi_{1}$, $\varphi_{2}$, $\varphi_{3}$, $\varphi_{4}$ from 0 to $2\pi$ as interval 0.3.}{\label{33}}
\end{center}
\end{figure}

{\sf The case of rank-6 states}~ We consider now the three-tangle
for a family of rank-6 mixed states: \be\label{rank6states}
\ba{lll}\varrho(t)=t|GHZ,2-\rangle\langle GHZ,2-|+(1-t)\sigma,\ea
\ee where \be\label{8} \ba{rcl}\sigma
&=&\frac{1}{11}|GHZ,1+\rangle\langle
GHZ,1+|+\frac{1}{11}|GHZ,1-\rangle\langle GHZ,1-|+
\frac{3}{11}|GHZ,2+\rangle\langle GHZ,2+|\\[3mm]
&&+\frac{3}{11}|GHZ,3+\rangle\langle GHZ,3+|+
\frac{3}{11}|GHZ,4+\rangle\langle GHZ,4+|.
\ea \ee
From the analysis for our rank-5 states, we know that $\sigma$
 has vanishing three-tangle, and the three-tangle for $\varrho(t)$ is given by
\be\label{imp} \ba{lll}\tau_{3}(\varrho(t))=\left\{
                                           \begin{array}{ll}
                                             0, & \hbox{$0\leq t\leq t_{0}$;} \\
                                             g_{I}(t), & \hbox{$t_{0}\leq t\leq t_{1}$;} \\
                                             g_{II}(t), & \hbox{$t_{1}\leq t\leq 1$,}
                                           \end{array}
                                         \right.
\ea \ee
where
$$\ba{rcl}
g_{I}(t)&=&t^2+\frac{6}{11}t(1-t)-\frac{27-24\sqrt{3}}{121}(1-t)^2-
\frac{24\sqrt{11}}{121}\sqrt{t(1-t)^3},\\[3mm]
g_{II}(t)&=&\frac{t-t_{1}}{1-t_{1}}+\frac{1-t}{1-t_{1}}g_{I}(t_{1}),~~~t_{0}=0.2143,~~~t_{1}=0.8290.
\ea
$$
We have the optimal decomposition \be\label{imp}
\ba{lll}\varrho(t)=\left\{
   \begin{array}{ll}
   \frac{t}{8t_{0}}\sum\Pi_{i}(t_0)+\frac{t_{0}-t}{t_{0}}\sigma, & \hbox{$0\leq t\leq t_{0}$;}\\[3mm]
   \frac{1}{8}\sum\Pi_{i}(t), & \hbox{$t_{0}\leq t\leq t_{1}$;} \\[3mm]
   \frac{1-t}{8(1-t_{1})}\sum\Pi_{i}(t_1)+
  \frac{t-t_{1}}{1-t_{1}}|GHZ,2-\rangle\langle GHZ,2-|, &
  \hbox{$t_{1}\leq t\leq 1$,}
                                           \end{array}
                                         \right.
\ea \ee where
\be\label{def2}\ba{l}\Pi_1(t)=|Z(t,0,0,0,0,0)\rangle\langle
Z(t,0,0,0,0,0)|,~~~
\Pi_2(t)=|Z(t,0,\pi,\pi,0,0)\rangle\langle Z(t,0,\pi,\pi,0,0)|,\\[3mm]
\Pi_3(t)=|Z(t,\pi,0,\pi,0,\pi)\rangle\langle
Z(t,\pi,0,\pi,0,\pi)|,~~~
\Pi_4(t)=|Z(t,\pi,\pi,0,0,\pi)\rangle\langle Z(t,\pi,\pi,0,0,\pi)|,\\[3mm]
\Pi_5(t)=|Z(t,\pi,0,\pi,\pi,0)\rangle\langle
Z(t,\pi,0,\pi,\pi,0)|,~~~
\Pi_6(t)=|Z(t,\pi,\pi,0,\pi,0)\rangle\langle
Z(t,\pi,\pi,0,\pi,0)|,\\[3mm]
\Pi_7(t)=|Z(t,0,0,0,\pi,\pi)\rangle\langle Z(t,0,0,0,\pi,\pi), ~~~
\Pi_8(t)=|Z(t,0,\pi,\pi,\pi,\pi)\rangle \langle
Z(t,0,\pi,\pi,\pi,\pi)|.\ea\ee
Obvious, all $\Pi_{i}(t_0)$ have vanishing three-tangle.

{\sf The case of rank-7 states} The three tangle of the following
rank-7 mixed states can be similarly calculated:
\be\label{rank7states} \ba{lll}\gamma(s)=s|GHZ,3-\rangle\langle
GHZ,3-|+(1-s){\zeta},\ea \ee where \be\label{rank7} \ba{rcl}
{\zeta}&=&\frac{1}{34}|GHZ,2-\rangle\langle GHZ,2-
|+\frac{3}{34}|GHZ,1+\rangle\langle
GHZ,1+|+\frac{3}{34}|GHZ,1-\rangle\langle GHZ,1- |\\[3mm]
&&+\frac{9}{34}|GHZ,2+\rangle\langle
GHZ,2+|+\frac{9}{34}|GHZ,3+\rangle\langle GHZ,3+|+
\frac{9}{34}|GHZ,4+\rangle\langle GHZ,4+|.
\ea \ee
Applying the similar approach above we get
\be\label{imp} \ba{lll}\tau_{3}(\gamma(s))=\left\{
\begin{array}{ll}
0, & \hbox{$0\leq s\leq s_{0}$,} \\
 g_{I}(s), & \hbox{$s_{0}\leq s\leq s_{1}$,} \\
 g_{II}(s), & \hbox{$s_{1}\leq s\leq 1$,}
  \end{array}
  \right.
\ea \ee
where
$$\ba{rcl}
g_{I}(s)&=&s^2+\frac{8}{17}s(1-s)-\frac{56-72\sqrt{3}}{289}(1-s)^2-
\frac{24\sqrt{102}}{289}\sqrt{s(1-s)^3},\\[3mm]
g_{II}(s)&=&\frac{s-s_{1}}{1-s_{1}}+\frac{1-s}{1-s_{1}}g_{I}(s_{1}),~~ s_{0}=0.2062,~ s_{1}=0.8375.
\ea
$$
We also can get the corresponding optimal decomposition for
$\gamma(s)$:
\be\label{imp}
\ba{lll}\gamma(s)=\left\{
   \begin{array}{ll}
   \frac{s}{8s_{0}}\sum\Pi_{i}(s_0)+\frac{s_{0}-s}{s_{0}}\zeta, & \hbox{$0\leq s\leq s_{0}$;}\\[3mm]
   \frac{1}{8}\sum\Pi_{i}(s), & \hbox{$s_{0}\leq s\leq s_{1}$;} \\[3mm]
   \frac{1-s}{8(1-s_{1})}\sum\Pi_{i}(s_1)+
  \frac{s-s_{1}}{1-s_{1}}|GHZ,3-\rangle\langle GHZ,3-|, &
  \hbox{$s_{1}\leq s\leq 1$,}
                                           \end{array}
                                         \right.
\ea \ee where
\be\label{def3}\ba{l}~~~\Pi_1(s)=|Z(s,0,0,0,0,0,0)\rangle\langle
Z(s,0,0,0,0,0,0)|,\\[3mm]
~~~\Pi_2(s)=|Z(s,0,\pi,0,\pi,\pi,\pi)\rangle\langle Z(s,0,\pi,0,\pi,\pi,\pi)|,\\[3mm]
~~~\Pi_3(s)=|Z(s,\pi,0,0,\pi,0,\pi)\rangle\langle
Z(s,\pi,0,0,\pi,0,\pi)|,\\[3mm]
~~~\Pi_4(s)=|Z(s,\pi,\pi,0,0,\pi,0)\rangle\langle Z(s,\pi,\pi,0,0,\pi,0)|,\\[3mm]
~~~\Pi_5(s)=|Z(s,0,\pi,\pi,0,0,\pi)\rangle\langle
Z(s,\pi,0,\pi,\pi,0,0,\pi)|,\\[3mm]
~~~\Pi_6(s)=|Z(s,0,0,\pi,\pi,\pi,0)\rangle\langle
Z(s,0,0,\pi,\pi,\pi,0)|,\\[3mm]
~~~\Pi_7(s)=|Z(s,\pi,\pi,\pi,\pi,0,0)\rangle\langle
Z(s,\pi,\pi,\pi,\pi,0,0)|,\\[3mm]
~~~\Pi_8(s)=|Z(s,\pi,0,\pi,0,\pi,\pi)\rangle \langle
Z(s,\pi,0,\pi,0,\pi,\pi)|.\ea\ee

{\sf The case of rank-8 states} The rank of a three-qubit mixed
state could be at most $8$. We now introduce a family of rank-8
mixed states:
$$\rho(r)=r|GHZ,4-\rangle\langle GHZ,4-|+(1-r){\eta},$$
where \be\label{rank-8}
\ba{rcl}{\eta}&=&\frac{1}{35}|GHZ,3-\rangle\langle
GHZ,3-|+\frac{1}{35}|GHZ,2-\rangle\langle
GHZ,2-|+\frac{3}{35}|GHZ,1+\rangle\langle GHZ,1+|\\[3mm]
&&+\frac{3}{35}|GHZ,1-\rangle \langle
GHZ,1-|+\frac{9}{35}|GHZ,2+\rangle\langle
GHZ,2+|\\[3mm]
&&+\frac{9}{35}|GHZ,3+\rangle\langle
GHZ,3+|+\frac{9}{35}|GHZ,4+\rangle\langle GHZ,4+|. \ea \ee
Obviously, $\tau_{3}({\eta})=0$. The three-tangle of $\rho(r)$ is
given by: \be\label{imp} \ba{lll}\tau_{3}(\rho(r))=\left\{
                                           \begin{array}{ll}
                                             0, & \hbox{$0\leq r\leq r_{0}$,} \\[2mm]
                                             g_{I}(r), & \hbox{$r_{0}\leq r\leq r_{1}$,} \\[2mm]
                                             g_{II}(r), & \hbox{$r_{1}\leq r\leq 1$,}
                                           \end{array}
                                         \right.
\ea \ee
where
$$
\ba{rcl}
g_{I}(r)&=&r^2+\frac{2}{5}r(1-r)-\frac{207-384\sqrt{3}}{1225}(1-r)^2
-\frac{128\sqrt{105}}{1225}\sqrt{r(1-r)^{3}},\\[3mm]
g_{II}(r)&=&\frac{r-r_{1}}{1-r_{1}}+\frac{1-r}{1-r_{1}}g_{I}(r_{1}),~~
r_{0}=0.2490,r_{1}=0.8649. \ea
$$
The optimal decomposition for $\rho(r)$ can be similarly obtained:
\be\label{imp}
\ba{lll}\rho(r)=\left\{
   \begin{array}{ll}
   \frac{r}{8r_{0}}\sum\Pi_{i}(r_0)+\frac{r_{0}-r}{r_{0}}\eta, & \hbox{$0\leq r\leq r_{0}$;}\\[3mm]
   \frac{1}{8}\sum\Pi_{i}(r), & \hbox{$r_{0}\leq r\leq r_{1}$;} \\[3mm]
   \frac{1-r}{8(1-r_{1})}\sum\Pi_{i}(r_1)+
  \frac{r-r_{1}}{1-r_{1}}|GHZ,4-\rangle\langle GHZ,4-|, &
  \hbox{$r_{1}\leq r\leq 1$,}
                                           \end{array}
                                         \right.
\ea \ee where
%\be\label{op-rank8-2} \ba{rcl} \rho(P)&=&\frac{1}{8}\sum\Pi_{i}(P),
%\ea \ee where
\be\label{def4}\ba{l}~~~\Pi_1(r)=|Z(r,0,0,0,0,0,0,0)\rangle\langle
Z(P,0,0,0,0,0,0,0)|,\\[3mm]
~~~\Pi_2(r)=|Z(r,0,0,0,\pi,\pi,\pi,\pi)\rangle\langle Z(r,0,0,0,\pi,\pi,\pi,\pi)|,\\[3mm]
~~~\Pi_3(r)=|Z(r,0,\pi,\pi,0,0,\pi,\pi)\rangle\langle
Z(r,0,\pi,\pi,0,0,\pi,\pi)|,\\[3mm]
~~~\Pi_4(r)=|Z(r,0,\pi,\pi,\pi,\pi,0,0)\rangle\langle Z(r,0,\pi,\pi,\pi,\pi,0,0)|,\\[3mm]
~~~\Pi_5(r)=|Z(r,\pi,0,\pi,0,\pi,0,\pi)\rangle\langle
Z(r,\pi,0,\pi,0,\pi,0,\pi)|,\\[3mm]
~~~\Pi_6(r)=|Z(r,\pi,0,\pi,\pi,0,\pi,0)\rangle\langle
Z(r,\pi,0,\pi,\pi,0,\pi,0)|,\\[3mm]
~~~\Pi_7(r)=|Z(r,\pi,\pi,0,0,\pi,\pi,0)\rangle\langle
Z(r,\pi,\pi,0,0,\pi,\pi,0),\\[3mm]
~~~\Pi_8(r)=|Z(r,\pi,\pi,0,\pi,0,0,\pi)\rangle \langle
Z(r,\pi,\pi,0,\pi,0,0,\pi)|.\ea\ee

\section{CKW inequality}

Given a family of mixed three-qubit states with the corresponding
three-tangle, one might check the CKW relations \cite{ckw-relation}.
For a pure three-qubit state $|\psi\rangle\in {\cal C}^2 \otimes
{\cal C}^2 \otimes {\cal C}^2$, with the reduced two-qubit density
matrices $\rho_{AB}=Tr_C(|\psi\rangle\langle\psi|)$,
$\rho_{AC}=Tr_B(|\psi\rangle\langle\psi|)$ and
$\rho_{A}=Tr_{BC}(|\psi\rangle\langle\psi|)$, one has the monogamy
relation $4det(\rho_{A})=C(\rho_{AB})^{2}+
C(\rho_{AC})^{2}+\tau_{3}(|\psi\rangle)$, where $C(\rho_{AB})$
(resp. $C(\rho_{AC})$) is the concurrence for the corresponding
reduced state $\rho_{AB}$ (resp. $\rho_{AC}$),
$\tau_{3}(|\psi\rangle)$ is the three-tangle of $|\psi\rangle$. For
mixed states, the following CKW inequality holds,
$4min[det(\rho_{A})]\geq C(\rho_{AB})^{2}+C(\rho_{AC})^{2}$. The CKW
inequality has been examined for the mixture of GHZ and W states in
\cite{rank2} and the mixture of GHZ, W and flipped-W states in
\cite{rank3}. In the following we check if the CKW inequality holds
for the states introduced in our paper.

As an example, we consider the case of rank-5 states. It is direct
to verify that these states satisfy
$C(\rho_{AB})^{2}+C(\rho_{AC})^{2}=0$. And the minimum one-tangle is
given by \be\label{9}
\ba{lll}4min[det(\rho_{A})]=1-\frac{8}{5}p(1-p)-\frac{9}{25}(1-p)^2+
\frac{6\sqrt{30}}{25}\sqrt{p(1-p)^{3}}. \ea \ee
%\begin{figure}[h]
%\begin{center}
%\resizebox{8cm}{!}{\includegraphics{ft.eps}}
%\caption{The\$p$-dependence of one-tangle (upper dotted line), sum
%of the squared concurrences (solid line along the horizontal axis)
%and three-tangle (solid line)}.{\label{33}}
%\end{center}
%\end{figure}
\begin{figure}[h]
\begin{center}
\resizebox{8cm}{!}{\includegraphics{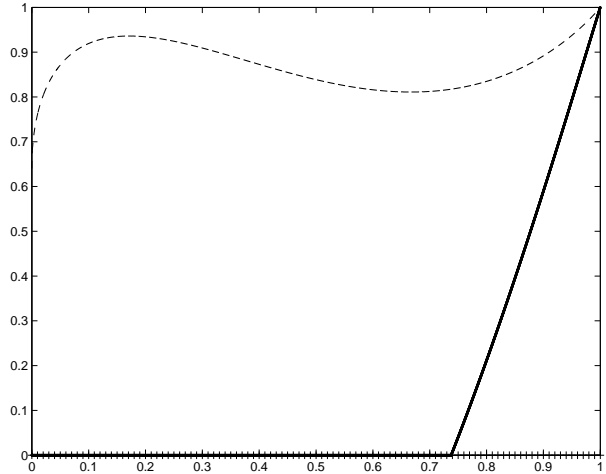}} \caption{The\ $p$
dependence of one-tangle (upper dotted line), sum of squared
concurrences (solid line along the horizontal axis) and three-tangle
(solid line).}{\label{33}}
\end{center}
\end{figure}
From Fig. 2 we see that the CKW inequality is obviously satisfied.
Moreover the inequality $4min[det(\rho_{A})]\geq C(\rho_{AB})^{2}+
C(\rho_{AC})^{2}+\tau_{3}(\psi)$ is also satisfied for these rank-5
states. In particular in the region $0\leq p \leq p_{0}=0.7377$,
both the concurrence and three-tangle are zero, but the one-tangle
is not zero.

\section{Conclusion}

We have constructed several classes of different ranked mixed states
in three-qubit system. We have provided explicit expressions for the
three-tangle and optimal decompositions for all these states. We
have also studied the relations between the CKW inequality and these
classes of states, and shown that the CKW inequality are satisfied for
these states. Concurrence of mixed two-qubit states has been applied to
study quantum phase transitions. It has been shown that the pairwise entanglement
of the nearest-neighbor two sites in spin-$1/2$ lattice models has special singularity at quantum
critical points \cite{pt}. It can be expected that multipartite
entanglement would reveal further relations between the quantum phase transitions and quantum entanglement.
Our results could help studies on applications of quantum entanglement
in all these related researches.

\vspace{0.8truecm}

\noindent {\bf Acknowledgments} The work is supported by NSFC (Grant
Nos.10871228, 10875081, 60873191, 60903152, 60821001), SRFDP (Grant
No. 200800131016, 20090005110010), Beijing Nova Program (Grant No.
2008B51), Key Project of Chinese Ministry of Education (Grant No.
109014), and Beijing Municipal Education Commission No.
KM200510028021 and No.KZ200810028013.

\end{document}